# Unusual temperature dependence of the band structure associated with local atomic distortion in monolayer $1T'$-WTe$_2$

Ryuichi Ando,[1] Katsuaki Sugawara,[1,2,3,*] Tappei Kawakami,[1] Koki Yanagizawa,[1] Ken Yaegashi,[1] Takashi Takahashi,[1] and Takafumi Sato [1,2,4,5,6,*]

[1]*Department of Physics, Graduate School of Science,*

*Tohoku University, Sendai 980-8578, Japan.*

[2]*Advanced Institute for Materials Research (WPI-AIMR),*

*Tohoku University, Sendai 980-8577, Japan.*

[3]*Precursory Research for Embryonic Science and Technology (PRESTO),*

*Japan Science and Technology Agency (JST), Tokyo 102-0076, Japan.*

[4]*Center for Science and Innovation in Spintronics (CSIS),*

*Tohoku University, Sendai 980-8577, Japan.*

[5]*International Center for Synchrotron Radiation Innovation Smart (SRIS),*

*Tohoku University, Sendai 980-8577, Japan and*

[6]*Mathematical Science Center for Co-creative Society (MathCCS),*

*Tohoku University, Sendai 980-8578, Japan.*

(Dated: November 29, 2024)

# Abstract

The ground state of monolayer 1$T'$-WTe$_2$ has been a target of intensive debate on whether or not it is a two-dimensional topological insulator (2D TI) associated with exciton formation. We investigated the band structure of an epitaxial monolayer 1$T'$-WTe$_2$ film grown on graphene/SiC(0001) in a wide temperature range of $T$ = 40 - 400 K by angle-resolved photoemission spectroscopy (ARPES). We observed an electron band above the Fermi level ($E_F$) slightly away from the Γ point, together with four hole bands below $E_F$ just at the Γ point. This signifies an indirect band gap exceeding 0.1 eV in support of the 2D-TI phase with the inverted band structure. We uncovered an unexpectedly large downward shift of valence bands upon cooling, accompanied with an upward shift of the conduction band. Comparison of the ARPES-derived band structure with first-principles band calculations suggests that the observed band shift is ascribed to the systematic local atomic distortion of tungsten atoms, which should be incorporated into the interpretation of unusual transport properties of 1$T'$-WTe$_2$.

**INTRODUCTION**

Searching for new functional two-dimensional (2D) materials is one of the central topics in materials science. Monolayer transition-metal dichalcogenides (TMDs) exhibit various exotic physical properties absent in the bulk counterpart [1], such as the Ising superconductivity [2–4], robust Mott-insulator phase [5, 6], and exciton Hall effect [7]. In particular, group-VI TMDs containing molybdenum (Mo) or tungsten (W) atoms have attracted much attention in connection with the topology, since they are predicted to be a 2D topological insulator (2D TI) with the helical edge state produced by the strong spin-orbit coupling (SOC) of Mo/W atoms. This was exemplified by the theoretical prediction for monolayer $1T'$-$MoS_2$ which hosts an inverted bulk-band structure consisting of the Mo $4d$ and S $3p$ bands [8] and subsequent experimental validation [9].

We focused on monolayer $1T'$-$WTe_2$ [its crystal structure is shown in Fig. 1(a)] among a few 2D-TI candidates of monolayer TMDs [10–13]. Bulk $WTe_2$ with $T_d$ structure is known to be a Weyl semimetal characterized by the surface Fermi-arc state [14–16], while it is predicted that the reduction of layer thickness down to the monolayer limit triggers a transition from a Weyl semimetal to a 2D-TI phase [8]. The 2D-TI phase of monolayer $1T'$-$WTe_2$ was experimentally supported by the transport and scanning tunneling microscopy (STM) measurements [12, 13] that signified the existence of metallic edge channels despite the bulk insulating nature [17]. Interestingly, monolayer $1T'$-$WTe_2$ shows superconductivity upon electron doping, making this system even more attractive in the viewpoint of topological superconductivity [18].

The 2D-TI nature of monolayer $1T'$-$WTe_2$ is attracting considerable attention in connection with the condensation of excitons. It was proposed that the strong coupling of conduction electrons and valence-band holes promotes the formation of excitons [19], leading to the exciton-induced 2D-TI phase [20, 21]. DFT (density-functional-theory) calculations with generalized gradient approximation (GGA) including SOC suggest a semimetallic band structure [8] with a negative band gap characterized by a small hole pocket at the $\Gamma$ point and a small electron pocket slightly away from the $\Gamma$ point (along the $\Gamma Y$ cut) separated from each other by a characteristic wave vector of $\mathbf{q}$. This behavior is inconsistent with the observed insulating behavior (i.e. positive band gap), and a plausible scenario to account for the positive band gap is the band folding characterized by the $\mathbf{q}$ vector and the direct hybridization between the conduction electron and valence hole bands to stabilize the excitonic gap [20, 21], leading to the transition from a band-inverted semimetal to a 2D-TI phase. The strong temperature dependence of conductivity and the absence of bulk conductance below ~100 K observed in exfoliated monolayer WTe$_2$ flakes [20, 21] are well interpreted in terms of the exciton-induced 2D-TI phase. However, to further clarify this scenario, the direct visualization of the band structure, in particular its temperature evolution, is useful, since the temperature-dependent spectroscopy measurements have played a crucial role in establishing the excitonic insulator phase in other systems [22–26].

In this letter, we report a temperature-dependent angle-resolved photoemission spectroscopy (ARPES) study of an epitaxial monolayer 1$T'$-WTe$_2$ film grown on graphene/SiC(0001). We observed a sizable temperature-dependent energy shift of the energy bands. We found that the spectroscopic energy gap persists at least up to $T = 400$ K, while the magnitude of band gap varies substantially with temperature. We have simulated the observed temperature-dependent band structure by DFT calculations with various lattice parameters, and found that the local distortion of W atoms is responsible for the unusual temperature evolution of band structure. We discuss implications of the present results in relation to other experiments and theories.

**EXPERIMENT AND CALCULATION**

Monolayer 1$T'$-WTe$_2$ film was grown on bilayer graphene by the molecular beam epitaxy (MBE) method [11, 26–28]. Bilayer graphene was fabricated by resistive heating of $n$-type 4H-SiC(0001) single-crystal wafer at 1100 °C for 15 min under high vacuum better than $1.0 \times 10^{-9}$ Torr. Subsequently, monolayer WTe$_2$ film was grown by evaporating W atoms on the bilayer graphene substrate in Te-rich atmosphere with keeping the substrate temperature at $T_s = 310$ °C. Then, the as-grown monolayer WTe$_2$ film was annealed at 310 °C for 30 min [28]. The reflection high-energy electron diffraction (RHEED) was used to monitor the *in-situ* growth of epitaxial films. After the growth, the film was transferred to the ARPES measurement chamber without exposing it to air. The RHEED pattern of the WTe$_2$ film shows a 2×1 streak pattern which confirms the monolayer nature of grown WTe$_2$ film. ARPES measurements were carried out using an MBS A-1 electron energy analyzer with a He discharge lamp constructed at Tohoku University. We used He I$\alpha$ photons with the energy of $hv = 21.218$ eV to excite photoelectrons. ARPES measurements were also performed at BL-28A in Photon Factory (PF) by using 60-eV photons with a micro beam spot of 12×10 $\mu$ m$^2$ [29]. The energy and angular resolutions were set to be 16-30 meV and 0.2°, respectively. The Fermi level ($E_F$) of samples was calibrated with a gold film deposited onto the substrate. First-principles band-structure calculations were carried out by using the QUANTUM-ESPRESSO package [30] with GGA [31]. To confirm the validity of the GGA calculation upon the lattice parameter variation, we have also carried out DFT calculations with HSE06 hybrid functional. The plane-wave cutoff energy and uniform $k$-point mesh were set to be 60 Ry and 10×5×1, respectively. The thickness of inserted vacuum layer was set to be more than 10 Å to prevent the interlayer interaction. The SOC was included in the calculations.

**RESULT AND DISCUSSION**

Figure 1(b) shows the plot of valence-band ARPES intensity at $T$ = 40 K for monolayer 1$T'$-WTe$_2$ measured along the ΓY cut in the rectangular Brillouin zone (BZ). One can recognize four hole bands centered at the Γ point sequentially marked from $E_F$ as H1-H4. The H1 band has a top of dispersion at the Γ point at the binding energy ($E_B$) of ~0.1 eV and does not cross $E_F$. An electron pocket slightly away from the Γ point along the ΓY cut predicted by the GGA calculation [see e.g. Fig. 3(b)] is absent, signifying a narrow-gap semiconductor nature with a band gap exceeding 0.1 eV at $T$ = 40 K. According to the DFT calculations [11, 32], the orbital character of H1-H4 bands is strongly **k**-dependent and complicated by the band inversion and the spin-orbit gap; the top of H1/H3 and H2 bands at the Γ point consist mainly of the W 5$d_{yz}$ and 5$d_{z^2}$ orbitals, respectively, whereas the H2 band has the W 5$d_{xy}$ and Te 6$p_y$ orbital character away from the Γ point. The overall agreement of the present ARPES result with the previous ARPES and DFT ones [11, 32] is in line with the topologically non-trivial ($Z_2$ = 1) character of monolayer 1$T'$-WTe$_2$ [17].

In the present study, we observed unexpectedly large temperature variation of the energy position of valence bands. For example, as shown in Fig. 1(c), the peak associated with the H4 band located at $E_B$ ~ 1.4 eV at $T$ = 40 K has moved toward $E_F$ at $T$ = 300 K. A similar behavior is also identified for the H2 band located at 0.5-0.6 eV. On the other hand, the shift of the H3 and H1 bands is not well visible while the spectral weight is significantly reduced at $T$ = 300 K similarly to the H2 and H4 bands. To see the temperature evolution of bands in more detail, we have carried out ARPES measurements with a finer temperature step with focusing on the near-$E_F$ energy range covering the H1 and H2 bands [Fig. 1(d)]. One can immediately recognize that the H2 band systematically moves toward $E_F$ on increasing temperature, while the change in the H1 band looks relatively small. We have confirmed the reproducibility of this temperature-dependent band shift by measuring several different samples under repeated thermal cycle and at different photon energies (see Fig. S1 in Supplemental Material [33]).

To gain further insight into the temperature evolution of band structure, we show in Fig. 2(a) ARPES-intensity plots near $E_F$ measured along the ΓY cut [purple line in Fig. 2(c)] at several temperatures of $T$ = 40 - 400 K. To visualize the states above $E_F$ by utilizing the finite population of Fermi-Dirac distribution (FD) function above $E_F$, we show in Fig. 2(b) the plot of ARPES intensity divided by the FD function at each temperature convoluted with the Gaussian resolution function. As immediately recognized in Fig. 2(a), upon decreasing temperature, the H2 band gradually shifts downward from $E_B$ ∼ 0.45 eV (at $T$ = 400 K) to ∼0.55 eV (at $T$ = 40 K) while keeping a similar shape of band dispersion, whereas the H1 band is nearly pinned at $E_B$ ∼ 0.1 eV. On the other hand, the ARPES intensity divided by the FD function in Fig. 2(b) signifies the existence of an electron-like conduction band (E1 band) above $E_F$ slightly away from the Γ point. The dispersion of E1 band becomes clearer at higher temperatures, especially above $T$ = 200 K. It is noted that while a similar electron band was previously reported in potassium-dosed monolayer 1$T'$WTe$_2$ [11], the visualization of bands above $E_F$ by utilizing the thermal broadening of the FD function would reflect the more intrinsic band structure of pristine sample because the K dosing may alter the band structure itself. This indicates that our sample is not situated in the degenerate-semiconductor regime, suggesting no obvious charge transfer from the graphene/SiC substrate and negligible self-doping from the defects in the crystal. It is noted that the faint intensity on both sides of the Γ point is not ascribed to a band dispersing below $E_F$, but it may originate from a tail of the bottom of E1 band located above $E_F$ associated with the lifetime broadening (see Fig. S2 in Supplemental Material [33]).

To discuss quantitatively the evolution of bands, we have mapped out the experimental band dispersion at each temperature by tracing the peak position in the energy distribution curve (EDC) [Fig. 2(d)]. One can again confirm the downward shift of the H2 band on cooling the sample.

Although the H1 band exhibits a similar trend, the magnitude of shift is much smaller. On the other hand, the E1 band slightly moves upward on decreasing temperature, as opposed to the downward shift of H1 and H2 bands, resulting in the enhancement of the size of band gap at lower temperatures. As shown in Fig. 2(e), the energy position of the top of H1 and H2 bands at the Γ point exhibits a linear temperature dependence in the temperature range of 40 - 400 K, with the total shift of 35 meV and 110 meV for the H1 and H2 bands, respectively. This again signifies a non-rigid-band-like behavior, as further supported by the opposite shift of the bottom of E1 band [red circles in Fig. 2(e)] that leads to the enhancement of band gap from 120 meV at $T$ = 400 K to 150 meV at $T$ = 200 K, as seen in Fig. 2(f). Obviously, these characteristics cannot be explained in terms of the temperature-dependent carrier doping from the graphene/SiC substrate, because the $E_F$ is located almost at the middle of the band gap irrespective of the temperature (in the range of 200-400 K). It is noted that a previous ARPES study with exfoliated monolayer $WTe_2$ encapsulated by graphene on $SiO_2$ substrate [34] shows that the top of the H1 band at the Γ point is located at $E_B$ = 0.07-0.08 eV and there exists tiny electron pocket from the E1 band at $T$ = 30-40 K. Although the previous and present ARPES results commonly show the positive nature of band gap, there exists a quantitative difference between the two, since the H1 top of MBE-grown $WTe_2$ is located at ~ 0.1 eV below $E_F$ and the electron band is unoccupied at $T$ = 40 K. This suggests that the $E_F$ position and the band-gap value are different between the MBE-grown and exfoliated samples. This difference may originate from the difference in the lattice strain and/or the electrostatic potential around the interface between two samples, presumably affecting the discussion of the exciton-induced 2D-TI phase.

Now that the temperature-dependent band evolution is established, the next step is to pin down the microscopic origin of such unexpected behavior. For this sake, we compare the band structure determined by ARPES with DFT calculations using various lattice parameters, since it is speculated that the change in the local lattice parameters may be responsible for the observed anomalous change in the band structure as in the case of Bi thin films [35]. We examined following three possibilities ; (i) in-plane lattice expansion/shrinkage, (ii) out-ofplane lattice expansion/shrinkage, and (iii) local atomic distortion that alters the internal coordinate of W atoms. The results of band calculations are shown in Fig. 3. It is known that the DFT calculation of monolayer $1T'$-$WTe_2$ with GGA shows a negative band gap, whereas the calculation with the hybrid functional (HSE06) reproduces a positive band gap [11, 32]. While it is ideal to utilize the DFT with HSE06, the calculation is time consuming so that it may not be so suitable for the systematic calculations with varying several different lattice parameters. Alternatively, we employed much faster GGA calculation because it is still useful to discuss semi-quantitatively the trend of experimental band dispersion. We have confirmed its validity by partially carrying out DFT calculations with HSE06, as detailed in Fig. S3 in Supplemental Material.

We show in Fig. 3(b) the calculated band structure for free-standing monolayer $1T'$WTe$_2$ along the ΓY cut for various in-plane ($a$-axis) lattice constants $d_a$ in Fig. 3(a). We found that the calculated E1 and H1 bands shift downward and upward, respectively, on increasing $d_a$ [see also the plot of the H1-H4-band top and the E1-band bottom against $d_a$ in Fig. 3(c)]. This behavior of the calculated E1 and H1 bands is consistent with the ARPES results showing the same trend on increasing temperature. On the other hand, the calculated H2 band moves downward as opposed to the upward shift of the calculated H1 band on increasing $d_a$, showing a clear distinction from the ARPES results where both the H1 and H2 bands move simultaneously toward the same direction [see Fig. 1(d) and Fig. 2(d)]. This suggests that the change in the in-plane lattice constant is unlikely to account for the experimental temperature evolution of bands. Next, we examined the possibility of the out-of-plane lattice parameter, i.e. the height difference between the top and bottom Te layers, $d_{Te}$ in Fig. 3(d). In this case, only the H3 band shows a strong $d_{Te}$ dependence [Figs. 3(e) and 3(f)], in contrast with the experimental observation [Fig. 1(c)] where the H2 and H4 bands show a larger shift than the H3 band. Thus, the second possibility (variation in $d_{Te}$) is ruled out. We then examined the third possibility associated with the height difference of W atoms in the $1T'$ structure [$d_W$, see Fig. 3(g)] and found that the variation of $d_W$ well captures the observed band shift. As seen in Fig. 3(h), the H1, H2, and H4 bands shift downward on increasing $d_W$ whereas the E1 band shifts upward, in qualitative agreement with the experimental band shift upon decreasing temperature. These features are also confirmed from the HSE06 based band calculations (see Fig. S3 in Supplemental Material [33]). As a result, the band gap is enhanced (although it is negative due to the limitation of GGA) on increasing $d_W$. It is plausible that the zigzag W layer becomes gradually flattened at higher temperatures, as seen in a stronger $1T'$-type (2×1) distortion at lower temperatures. We have also examined some other lattice parameters in the DFT calculations such as the $b$-axis lattice constant (see Fig. S4 in Supplemental Material [33]), and found that none of them reproduce the experimental band behavior. We thus concluded that $d_W$ is a key parameter to govern the temperature-dependent band shift.

Now we discuss the implication of present results in relation to the proposed excitoninduced 2D TI phase [20, 21]. We discuss this possibility in terms of (i) the exciton binding energy and (ii) the charge-density wave (CDW) associated with the exciton condensation. Regarding (i), we estimate the experimental band gap to be ∼150 meV at room temperature. If excitons condensate, this band gap corresponds to the exciton binding energy. On the other hand, the exciton binding energy in monolayer $1T'$-WTe$_2$ predicted by the GGA calculation is ∼330 meV [21], more than twice larger than the experimental band gap. Since the band gap is underestimated in the GGA calculation as discussed above, for a better estimation of the exciton binding energy, it would be necessary to refer to the DFT calculation with HSE06 hybrid function that properly reproduces the experimental band gap of ∼ 150 meV (at room temperature). Although the exciton binding energy estimated from HSE06 is not available at the moment, it is expected to be smaller than 330 meV for the $1T'$-WTe$_2$/graphene/SiC system, because the observed larger band gap would likely reduce the energy gain associated with the exciton formation. Nevertheless, when considering the fact that the experimental band-gap value (∼ 150 meV) coincides well with that predicted by HSE06 without incorporating the exciton formation [11, 36, 37], it is suggested that an additional enhancement of the band gap due to the exciton formation may cause a stronger deviation of the band-gap values between the calculation and experiment. This seems to be not so compatible with the exciton formation scenario. Regarding (ii), since the stabilization of excitons generally requires a direct band gap, the band folding with a characteristic wave vector **q** corresponding to the **k** difference between the E1-band bottom and the H1band top is expected to occur to cause the formation of CDW, as commonly recognized in other excitonic insulator candidates such as $1T$-TiSe$_2$ [22–26]. In this case, the ARPES intensity associated with the band folding would be recognized [22, 26, 38, 39], as reported in some TMDs with exciton formation (see section 5 in Supplemental Material [33]). On the other hand, as seen in Fig. 2(a), we observe no obvious signature for such a band folding or anomalies associated with the band hybridization between the folded and main E1/H1 bands, in line with the micro-ARPES study with an exfoliated flake sample [34]. Moreover, the simple linear temperature dependence of the band shift and the band gap [Figs. 2(e) and 2(f)] is not compatible with the formation of excitonic gap (CDW gap). These results suggest that our MBE-grown monolayer $1T'$-WTe$_2$ film is characterized by the normal 2D TI phase without exciton condensation in a wide temperature range at least up to 400 K. At the moment, it is unclear how our result reconciles with the transport measurements with a flake sample that supports the exciton-induced 2D-TI phase. To clarify this point, it would be necessary to carry out temperature-dependent ARPES measurements with an exfoliated flake sample and directly compare the band evolution with that of a MBE-grown film.

Finally, we discuss the implication of unusual temperature-dependent band evolution. It is generally expected that the lattice expands on heating (i.e. positive thermal expansion coefficient) due to the thermal vibration of atoms, as reported in a Bi(111) thin film on Si(111) [35] where the interlayer spacing expands on heating. On the other hand, in monolayer $1T'$WTe$_2$, the W-W bond length expands on decreasing temperature due to the increase in $d_W$. This can be viewed as a negative thermal expansion when neglecting the contribution from Te layers. Interestingly, the negative thermal expansion was recently identified in layered bulk crystals of Ta$_2$NiSe$_5$ [40] and $T_d$-MoTe$_2$ [41], although it is unclear whether they share a common mechanism. Here, we point out a possibility that the temperature-dependent band shift is associated with the topological nature of monolayer $1T'$-WTe$_2$, although it may be too early to speculate along this line. In contrast to the normal band structure, the inverted band structure is likely to cause the charge redistribution between W and Te atoms upon temperature variation, leading to the change in the local bond length and the enhancement of atomic displacement. To address this issue, it is necessary to carry out ARPES and DFT investigations on other atomic-layer TMDs with both topological and non-topological characteristics.

## SAMMARY

The present temperature-dependent ARPES study on monolayer $1T'$-WTe$_2$ grown epitaxially on graphene/SiC(0001) has revealed an unexpectedly large variation of the band structure with temperature. The observed temperature-dependent band modulation is qualitatively reproduced by the DFT calculations that incorporate the systematic variation in the height difference between two adjacent W atoms. We propose that the temperaturedependent change in the lattice parameter needs to be incorporated to properly discuss the transport properties associated with the excitonic and 2D-TI phase in $1T'$-WTe$_2$.

## ACKNOWLEDGMENTS

This work was supported by JST-CREST (no. JPMJCR18T1), JST-PRESTO (no. JPMJPR20A8), Grant-in-Aid for Scientific Research (JSPS KAKENHI Grant Numbers

JP18H01821, JP20H01847, JP20H04624, JP21H01757, JP21K18888, JP21H04435, and JP22J13724, and JP24H01166), KEK-PF (Proposal No. 2020G669, 2021S2-001, 2022G007, and 2024G002), Foundation for Promotion of Material Science and Technology of Japan, Samco Foundation, and World Premier International Research Center, Advanced Institute for Materials Research. T. K., K. Yanagizawa, and K. Yaegashi acknowledge support from GP-Spin at Tohoku University. T. K. and K. Yanagizawa also acknowledges support from JSPS.

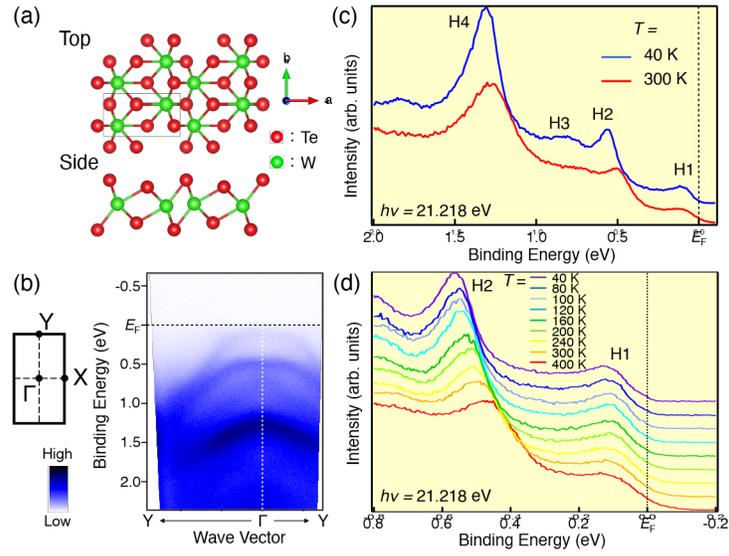

FIG. 1. (a) Top and side views of crystal structure of monolayer 1$T'$-WTe$_2$. (b) (left) First BZ of 1$T'$-WTe$_2$ and (right) ARPES-intensity plot for monolayer 1$T'$-WTe$_2$ measured with the He-I$\alpha$ resonance line along the $\Gamma$Y cut at $T = 40$ K. (c) EDCs at the $\Gamma$ point measured at $T = 40$ and 300 K. (d) Temperature dependence of EDC near $E_F$ at the $\Gamma$ point.

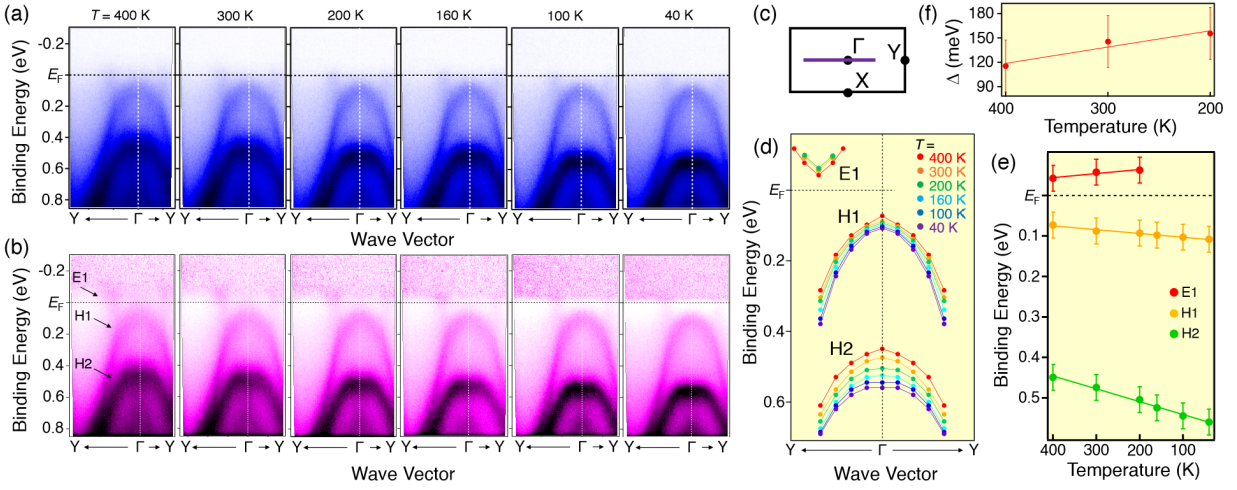

FIG. 2. (a) Temperature dependence of ARPES intensity for monolayer 1T'-WTe$_2$ measured along the ΓY cut. (b) Same as (a) but divided by the FD function at each temperature convoluted with the resolution function. (c) First BZ of 1T'-WTe$_2$ together with the **k** cut (purple line) where the intensity plot in (a) and (b) was obtained. (d) Plot of experimental band dispersion for E1, H1, and H2 bands at various temperatures obtained by the numerical fitting of EDCs divided by the FD function. (e) Temperature dependence of the band energy for the bottom of E1 band and the top of H1 and H2 bands. (f) Temperature dependence of experimental band gap estimated from the energy difference in the peak position of EDCs between the valence-band top at the Γ point of the H1 band and the conduction-band bottom slightly away from the Γ point of the E1 band.

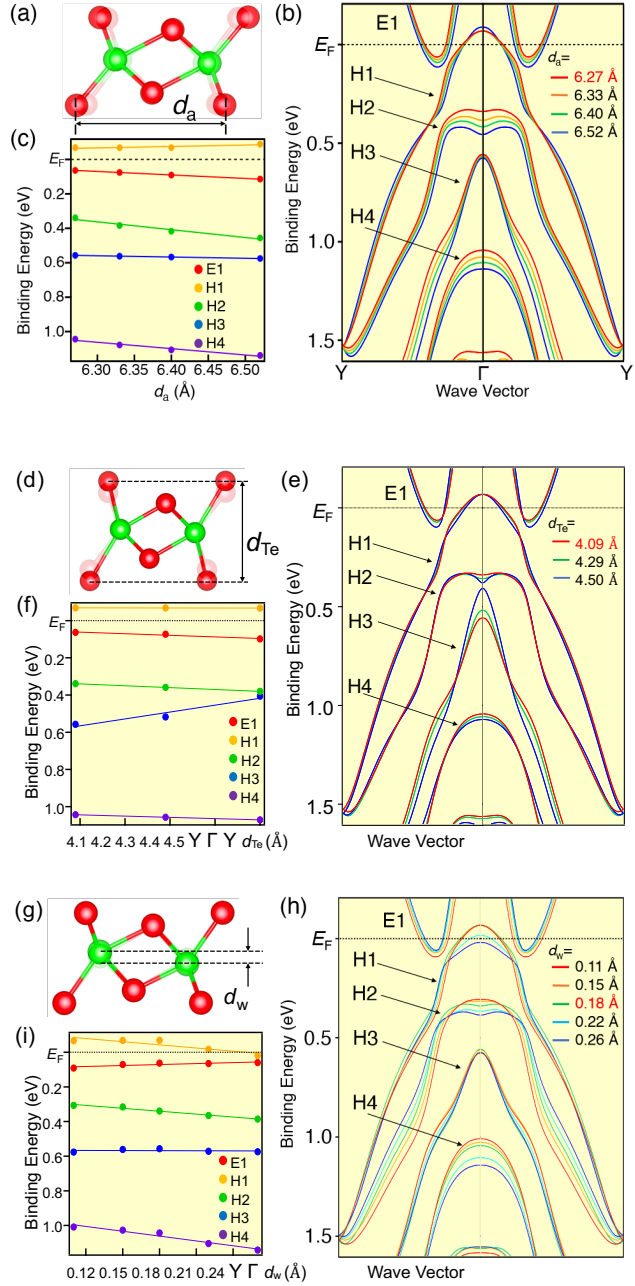

FIG. 3. (a) Schematics of in-plane uniaxial lattice expansion in 1$T'$-WTe$_2$. $d_a$ corresponds to the $a$-axis lattice constant. (b) Calculated band structure for monolayer 1$T'$-WTe$_2$ with various $d_a$'s. (c) $d_a$ dependence of the energy position of the bottom of E1 band and the top of H1 and H2 bands in (b). (d)-(f) Same as (a)-(c), but for out-of-plane lattice expansion. $d_{Te}$ is the height difference between the top and bottom Te atoms. (g)-(i) Same as (a)-(c), but the local displacement of W atoms with fixing the position of Te atoms is taken into account. $d_W$ is the height difference between two W atoms in the unit cell.

# Supplemental Material:

# "Unusual temperature dependence of the band structure associated with local atomic distortion in monolayer 1$T$'-WTe$_2$"

Ryuichi Ando,[1] Katsuaki Sugawara,[1,2,3] Tappei Kawakami,[1] Koki Yanagizawa,[1] Ken Yaegashi,[1] Takashi Takahashi,[1] and Takafumi Sato[1,2,4,5,6]

[1] *Department of Physics, Graduate School of Science, Tohoku University, Sendai 980-8578, Japan.*

[2] *Advanced Institute for Materials Research (WPI-AIMR), Tohoku University, Sendai 980-8577, Japan.*

[3] *Precursory Research for Embryonic Science and Technology (PRESTO), Japan Science and Technology Agency (JST), Tokyo 102-0076, Japan.*

[4] *Center for Science and Innovation in Spintronics (CSIS), Tohoku University, Sendai 980-8577, Japan.*

[5] *International Center for Synchrotron Radiation Innovation Smart (SRIS), Tohoku University, Sendai 980-8577, Japan.*

[6] *Mathematical Science Center for Co-creative Society (MathCCS), Tohoku University, Sendai 980-8578, Japan.*

**S1. Reproducibility of the temperature dependent band shift**

To check the reproducibility of the temperature dependent ARPES data, we have carried out ARPES measurements with synchrotron radiation besides a laboratory-based photon source [He-Iα line ($h\nu$ = 21.218 eV)]. Figure S1 shows the temperature dependence of EDC at the Γ point obtained with $h\nu = 60$ eV for monolayer 1$T$'-WTe$_2$ fabricated with the same growth condition as that of the laboratory-based system. One can recognize that the H2, H3, and H4 bands monotonically shift closer to the Fermi level ($E_F$) on increasing temperature, reproducing the result obtained with the He-Iα line presented in Fig. 1.

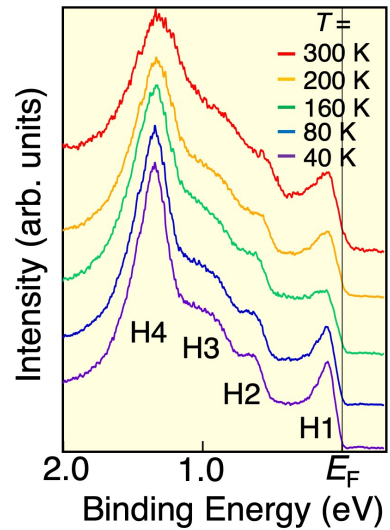

**Fig. S1**: Temperature dependence of EDC at the Γ point obtained with $h\nu$ = 60 eV.

### S2. Origin of the finite intensity on both sides of the Γ point below $E_F$

Figures S2(a) and S2(b) show a series of raw EDCs along the ΓY cut obtained at $T = 40$ K and 300 K, respectively. In the region indicated by blue dashed circles, one can see a faint spectral weight near $E_F$ at both $T = 40$ K and 300 K. Unlike the case of dispersive holelike valence band centered at the Γ point, it is hard to recognize the energy dispersion for this feature. Considering the observed monotonous spectral-weight reduction on moving away from the E1-peak position as seen in Fig. 2(b), this feature likely originates from the tail of conduction-band bottom located above $E_F$ associated with the lifetime broadening.

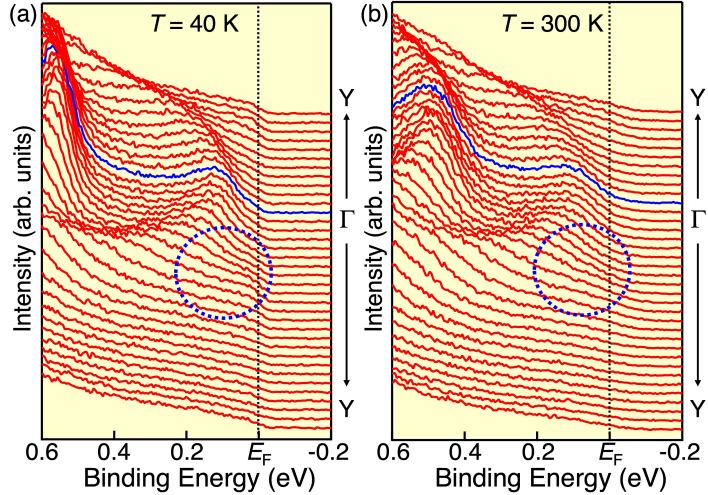

**Fig. S2:** (a), (b) Series of EDCs around the Γ point measured at $T = 40$ and 300 K, respectively.

### S3. DFT calculations with HSE06

To justify that the GGA calculation which underestimates the band gap can be used to discuss the observed temperature-dependent band shift, we have made a direct comparison of the calculated band structure between GGA and HSE06 hybrid functional by selecting a few representative lattice parameters. In both GGA and HSE calculations, we at first fully relaxed the crystal structure and estimated the optimized $d_W$ value (height difference between two W atoms in the unit cell) to be 0.18 and 0.20 Å for GGA and HSE, respectively, while the in-plane lattice constants are $a$ = 6.28Å and $b$ = 3.50 Å for both cases. Then, we systematically changed the $d_W$ value to directly compare the energy shift of the E1 and H1-H4 bands. Figures S3(a) and S3(b) show a side-by-side comparison of the calculated band structure between GGA and HSE for five different $d_W$ values. It is evident that the general trend of the band movement upon changing $d_W$, e.g. a downward shift of H1, H2, H3, and H4 as well as an upward shift of E1 with increasing $d_W$ in the GGA calculation is qualitatively reproduced in the HSE calculation despite the quantitative difference in the band-gap value between the two calculations (note that the energy shift of H3 band is lager in the HSE calculation, but our argument in the main text is essentially not affected by this difference). This suggests that the discussion on the band shift based with the GGA calculation works well to pin down the origin of observed temperature-dependent band shift. This may be reasonable because the GGA and HSE calculations commonly predict the inverted band character although the band-gap size is quantitatively different between the two calculations.

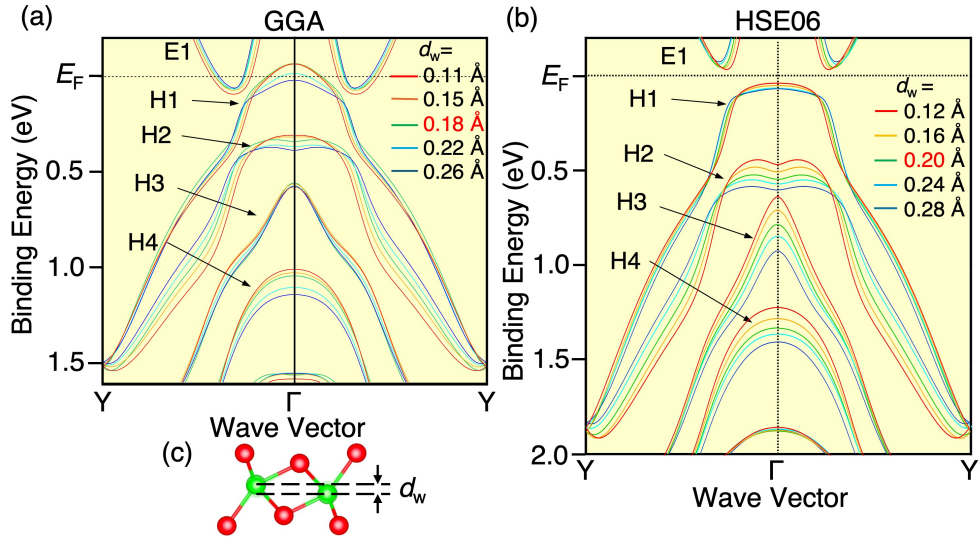

**Fig. S3**. (a), (b) Comparison of calculated band structure along the ΓY cut for representative $d_W$ values between GGA and HSE06, respectively. $d_W$ is the height difference between two W atoms in the unit cell as highlighted in (c). (c) Schematics of the side view of $1T'$-WTe$_2$.

## S4. Band calculations with various lattice parameters

Figures S4(b) and S4(c) show the band structure calculated by varying the $b$-axis lattice constant and by simultaneously varying the $a$- and $b$-axis lattice constants with keeping the same $a/b$ ratio, respectively. In both cases, the H1 and H2 bands shift toward opposite directions from each other, inconsistent with the ARPES result shown in Fig. 2.

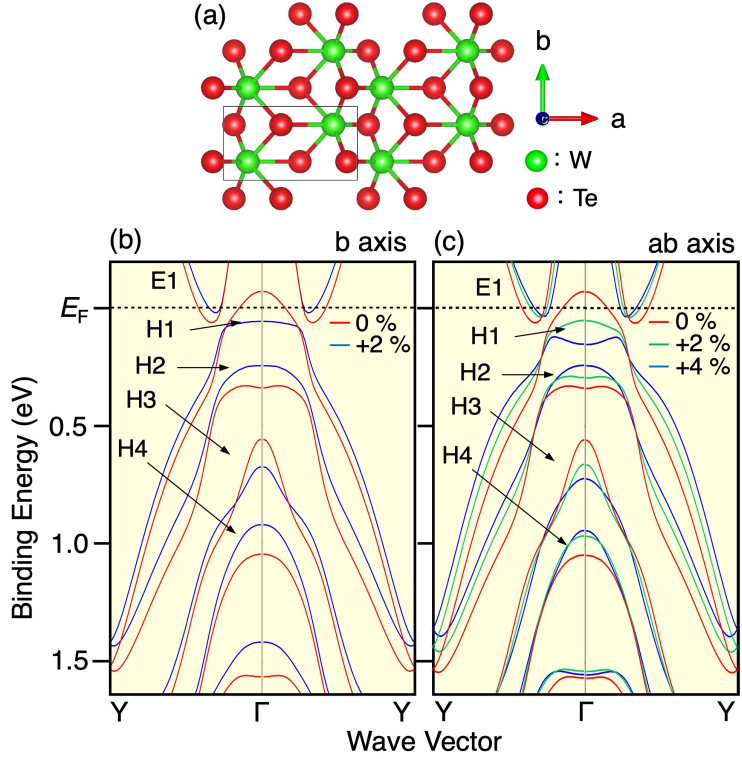

**Fig. S4:** (a) Top view of crystal structure of monolayer 1$T'$-WTe$_2$. (b), (c) Calculated band structures for monolayer 1$T'$-WTe$_2$ simulated (b) by varying the $b$-axis length and (c) by simultaneously varying the $a$- and $b$-axis lengths with keeping the $a/b$ ratio constant.

This also supports that the height difference between two W atoms ($d_W$) is a key parameter to cause the large temperature variation of band energy.

**S5. Comparison of band folding associated with CDW among TMDs**

To examine the possibility of exciton formation, it would be useful to examine to what extent the formation of excitons is sensitive to the backfolding of bands in the ARPES data. For this sake, we refer to the established excitonic insulator materials and examine the experimental intensity of backfolded bands in the CDW phase. As summarized in Table S1, in some monolayer TMDs such as TiSe$_2$, ZrTe$_2$, and HfTe$_2$ with the exciton formation temperature ($T_{ex}$; same as the CDW transition temperature) of ~74-230 K, the spectral weight of backfolded bands $I_{folded}$ estimated from the ARPES intensity is comparable to or even stronger than that of main bands. While this could be due to the high $T_{ex}$ value and/or the commensurate nature of CDW periodicity, it is inferred that the intensity of backfolded bands is generally well recognizable in other TMDs. In 1$T'$-WTe$_2$, the insulating gap already opens at $T$ = 400 K and the band gap smoothly evolves on decreasing temperature without any anomalies, as shown in Figs. 2(b) and 2(d) of the main text. If the observed gap is associated with the exciton formation, it is inferred that $T_{ex}$ of our 1$T$'-WTe$_2$ film would be well above those for other TMDs shown in Table S1. However, we found from the ARPES intensity at the expected backfolding wave vector (at the wave vector where the E1-band bottom is located) that the intensity of backfolded band, if it exists, is well below 10% of that of the main band, by analyzing the spectral intensity at $T$ = 40 K in Fig. 2(a). To exclude the possibility of weak band folding originating from an artifact associated with photoionization cross-section or photoelectron matrix-element effect, we also carried out ARPES measurements at different photon energies in synchrotron facility (e.g $h\nu$ = 60 eV) and found that the backfolded band is always absent. These arguments, together with the experimental fact that previous scanning tunneling microscopy (STM) studies have not detected a superlattice modulation ascribable to the occurrence of CDW, suggest that excitons are unlikely formed in our epitaxial monolayer 1$T$'-WTe$_2$ film.

| Material | Bulk/monolayer | $T_{ex}$ | Periodicity | $I_{folded}$ at M point | References |
|---|---|---|---|---|---|
| TiSe$_2$ | Monolayer | 200~230 K | 2 x 2 | ~2 | K. Sugawara *et al.*, ACS Nano **10**, 1341 (2016). |
| TiSe$_2$ | Bulk | ~200 K | 2 x 2 x 2 | ~6 | H. Cercellier *et al.*, Phys. Rev. Lett. **99**, 146403 (2007). |
| ZrTe$_2$ | Monolayer | 150~180 K | 2 x 2 | ~5 | Q. Gao *et al.*, Nat. Commun. **14**, 994 (2023). |
| HfTe$_2$ | Monolayer | 74-175K | 2 x 2 | ~1 | Q. Gao *et al.*, Nat. Phys. **20**, 597 (2024) |

Table S1: I Intensity of backfolded band upon exciton formation ($I_{folded}$) with respect to that of main band in some TMDs, estimated from the intensity at the M point of hexagonal Brillouin zone at low temperatures. $T_{ex}$ represents the exciton condensation temperature (same as the CDW transition temperature).